
\magnification\magstep1

\openup 3\jot

\input mssymb
\def\hbar{\mathchar '26\mkern -9muh}

\def\half{{\textstyle {1 \over 2}}}

\def\twothirds{{\textstyle {2 \over 3}}}

\def\pmb#1{\setbox0=\hbox{#1} \kern-.025em\copy0\kern-\wd0
\kern0.05em\copy0\kern-\wd0 \kern-.025em\raise.0433em\box0 }
\def\pmbh#1{\setbox0=\hbox{#1} \kern-.12em\copy0\kern-\wd0
\kern.12em\copy0\kern-\wd0\box0}
\def\sqr#1#2{{\vcenter{\vbox{\hrule height.#2pt \hbox{\vrule
width.#2pt height#1pt \kern#1pt \vrule width.#2pt} \hrule
height.#2pt}}}}

\def\rchi{{\raise 2pt \hbox {$\chi$}}}
\def\rga{{\raise 2pt \hbox {$\gamma$}}}

\def\({\left(}
\def\){\right)}
\def\<{\left\langle}
\def\>{\right\rangle}

\def\[{\left[}
\def\]{\right]}
\let\text=\hbox
\def\pt{\partial}
\def\eps{\epsilon}
\def\kap{\kappa}

\def\ol{\overline}
\def\ul{\underline}

\def\de{\delta}
\def\lam{\lambda}

\def\ti{\tilde}

\def\cl{\centerline}

\def\ni{\noindent}
\def\wti{\widetilde}

\null
\vskip 1 true in
\cl {\bf Physical states in $ N = 1 $ supergravity}
\vskip 1 true in
\cl {P.D. D'Eath}
\cl {D.A.M.T.P., University of Cambridge, Silver Street,}
\cl {Cambridge CB3 9EW, U.K.}
\vskip 1.5 true in
\cl {\bf Abstract}

\noindent By solving the supersymmetry constraints for physical
wave-functions, it is shown that the only two allowed bosonic states
in $ N = 1 $ supergravity are of the form const.~exp~$(\pm I / \hbar)
$, where $ I $ is an action functional of the three-metric. States
containing a finite number of fermions are forbidden. In the case that
the spatial topology is $ S^3 $, the state const.~exp~$(- I / \hbar)$
is the wormhole ground state, and the state const.~exp~$(I / \hbar)$
is the Hartle--Hawking state. $ N = 1 $ supergravity has no quantum
ultraviolet divergences, and no quantum corrections.
\vskip 20 pt
\centerline {PACS numbers 04.60. + n, 04.65. + e, 98.80 Dr.}

\vfill\eject

Usually, quantum supergravity is treated in terms of scattering theory
using a path-integral approach. However, the full quantum theory can
instead be treated non-\break perturbatively by studying the quantum
constraints acting on the wave function [1], i.e.~by taking an
approach based on functional differential equations. This approach is
used here to show that there are only two purely bosonic
wave-functions in $ N = 1 $ supergravity, which have the simple form $
\exp ( \pm I / \hbar ) $, where $ I $ is an action functional of the
3-metric. Further, states containing a finite number of fermions are
forbidden.

A wave-function can be taken to be of the form $ \Psi \( e^{A
A'}_{~~~~i} (x),~\psi^A_{~~i} (x) \) $. Here, using 2-component
notation [1], $ e^{A A'}_{~~~~i} (x) $ is the spatial tetrad, which
gives the 3-metric as $ h_{i j} = - e^{A A'}_{~~~~i} e_{A A'j} $, and
$ \( \psi^A_{~~i},~\wti
\psi^{A'}_{~~i} \) $ is the spatial gravitino field, taken to be an
odd Grassmann quantity. The wave-function can equivalently be
described by $ \wti \Psi \( e^{A A'}_{~~~~i} (x),~\wti \psi^{A'}_{~~i}
(x) \) $, which is related to $
\Psi $ by a fermionic Fourier transform [1]. A physical wave-function
must obey the Lorentz and supersymmetry constraints $$ \eqalignno {
J^{A B} \Psi &= 0~, \ \ \ol J^{A' B'} \Psi = 0~, &(1),(2) \cr S^A \Psi
&= 0, \ \ \ \ \ \ol S^{A'} \Psi = 0~. &(3),(4) \cr } $$ The remaining
Hamiltonian constraints $ {\cal H}_{A A'} \Psi = 0 $, suitably
ordered, will be implied by the above constraints [2]. The Lorentz
constraints imply that $ \Psi $ is invariant under local Lorentz
rotations applied to the arguments $ \( e^{A A'}_{~~~~i}
(x),~\psi^A_{~~i} (x) \) $.  Thus all Lorentz indices must be
contracted together in $ \Psi $. Hence $
\Psi $ can only contain an even number $ \psi^0, \psi^2, \psi^4,
\ldots $ of fermions. Further the supersymmetry constraints do not mix
fermion number; thus one can study the states proportional to $
\psi^0, \psi^2, \psi^4,
\ldots $ separately. The constraint $ \ol S_{A'} \Psi = 0 $ reads $$
\eps^{i j k} e_{A A' i}\( \; ^{3 s} D_j \psi^A_{~~k} \) \Psi - \half
\hbar
\kap^2
\psi^A_{~~i} {\de \Psi \over \de e^{A A'}_{~~~~i}} = 0~. \eqno (5) $$
Here $ ^{3 s} D_j $ is the 3-dimensional covariant derivative on
spinors without torsion [1], and $ \kap^2 = 8 \pi $. The constraint $
S_A \Psi = 0 $ reads $$ ^{3 s} D_i \( {\de \Psi \over \de
\psi^A_{~~i}} \) + \half \hbar
\kap^2 {\de
\over \de e^{A A'}_{~~~~i}} \( D^{B A'}_{~~~~j i} {\de \Psi \over \de
\psi^B_{~~j}} \) = 0~, \eqno (6) $$ where $$ D^{B A'}_{~~~~j i} = - 2
i h^{- {1 \over 2}} e^{B B'}_{~~~~i} e_{C B' j} n^{C A'}~. \eqno (7)
$$ Here $ h = \det \( h_{i j} \) $, and $ n^{A A'} $ is the spinor
version of the unit future-pointing normal $ n^\mu $ to a surface $
x^0 = {\rm constant} $. This is defined as a function of the $ e^{A
A'}_{~~~~i} $ by $$ n^{A A'} e_{A A' i} = 0~, \ \ \ \ n^{A A'} n_{A
A'} = 1~. \eqno (8) $$ The $ S_A $ constraint is more easily
understood in the representation $ \wti
\Psi \( e^{A A'}_{~~~~i}, \wti \psi^{A'}_{~~i} \) $, where it reads $$
\eps^{i j k} e_{A A' i} \(\;^{3 s} D_j \wti \psi^{A'}_{~~k} \) \wti
\Psi + \half \hbar
\kap^2 \wti \psi^{A'}_{~~i} {\de \wti \Psi \over \de e^{A A'}_{~~~~i}}
= 0~. \eqno (9) $$

We now restrict attention to a purely bosonic wave function $ \Psi \(
e^{A A'}_{~~~~i} (x) \) $. This automatically obeys the constraint $
S_A \Psi = 0 $ [Eq.~(6)]. Consider then the constraint $ \ol S_{A'}
\Psi = 0 $ [Eq.~(5)].  Since $ \Psi $ is purely bosonic, one can
rewrite Eq.~(5) as $$ \eps^{i j k} e_{A A' i} \(\;^{3 s} D_j
\psi^A_{~~k} \) - \half \hbar \kap^2
\psi^A_{~~i} {\de (\ln \Psi ) \over \de e^{A A'}_{~~~~i}} = 0~. \eqno
(10) $$ One now remarks that this gives a unique expression for $ \de
\( \ln \Psi \) / \de e^{A A'}_{~~~~i} (x) $ as a functional of $ e^{A
A'}_{~~~~i} (x) $. To see this, note that the field $ \psi^A_{~~i} (x)
$, equivalently given by $$ \psi^A_{~~B B'} = \psi^A_{~~i} e_{B
B'}^{~~~~i}~, \eqno (11) $$ can be decomposed into a spin - ${3 \over
2} $ part $ \rga_{A B C} $ and a spin - $ \half $ part $ \beta_A $ as
[1] $$ \psi_{A B B'} = - 2 n^C_{~~B'} \rga_{A B C} + \twothirds \(
\beta_A n_{B B'} + \beta_B n_{A B'} \) - 2 \eps_{A B} n^C_{~~B'}
\beta_C~, \eqno (12) $$ where $ \rga_{A B C} = \rga_{(A B C)} $ is
totally symmetric. One can then decompose $ \psi^A_{~~i} (x) $ into
spinor harmonics, using the bases $
\rga^{(n)}_{A B C} $ and $ \beta^{(m)}_A $ which obey $$ \eqalignno {
e^{A A' j} \;^{3 s} D_j \rga^{(n)}_{A B C} &= i \mu_n n^{A A'}
\rga^{(n)}_{A B C}~, &(13) \cr e^{A A' j} \;^{3 s} D_j \beta^{(m)} &=
i \lam_m n^{A A'} \beta^{(m)}_A~.  &(14) \cr } $$ [The indices $ m , n
$ have been written as discrete, for a compact 3-manifold, and should
be replaced by continuous indices for a non-compact manifold.] One
writes $$ \eqalignno {
\rga_{A B C} &= \sum_n c_n \rga^{(n)}_{A B C}~, &(15) \cr
\beta_A &= \sum_m b_m \beta^{(m)}_A~, &(16) \cr } $$ where the
coefficients $ c_n $ and $ b_m $ are odd Grassmann quantities, and $
\rga^{(n)}_{A B C},~\beta^{(m)}_A $ are even. One substitutes
Eqs.~(15, 16) into Eq.~(12), to obtain an expansion for $$
\psi_A^{~~i} = - \psi_A^{~~B B'} e_{B B'}^{~~~~i}~. \eqno (17) $$ The
first term in the constraint (10) can then be expanded, using
Eqs.~(13, 14), as $$ \eps^{i j k} e_{A A' i} \(\;^{3 s} D_j
\psi^A_{~~k} \) = - {8 \over 3} h^{1 \over 2} n^A_{~~A'} \sum_m \lam_m
b_m \beta^{(m)}_A~. \eqno (18) $$

Since the constraint (10) must hold for all choices of $ \psi^A_{~~i}
(x) $, it must equivalently hold for all choices of the coefficients $
b_m $ and $ c_n $ in Eqs.~(15, 16). Hence $$ e_{B B'}^{~~~~i} n^{C B'}
\rga^{(n) A B}_{~~~~~~C} {\de ( \ln \Psi) \over
\de e^{A A'}_{~~~~i}} = 0~, \eqno (19) $$ $$ \eqalignno { &- {8 \over
3} h^{1 \over 2} \lam_m n^A_{~~A'} \beta^{(m)}_A
\cr &+ \half \hbar
\kap^2 e_{B B'}^{~~~~i} \( \twothirds \beta^{(m) B} n^{A B'} - 2
\eps^{A B} n^{C B'} \beta^{(m)}_C \) {\de (\ln \Psi) \over \de e^{A
A'}_{~~~~i}} = 0~, &(20) \cr } $$ for all $ m $ and $ n $. These are
the same equations as would appear if $
\psi^A_{~~i} (x) $ were taken to be bosonic in Eq.~(10). One can then
contract Eqs.~(19, 20) with an arbitrary bosonic field $$ \de_{A'} =
\sum_p d_p \de^{(p)}_{A'}~, \eqno (21) $$ where $$ e^{A
A'}_{~~~~j}\;^{3 s} D_j \de^{(p)}_{A'} = - i \lam_p n^{A A'}
\de^{(p)}_{A'}~. \eqno (22) $$ Eqs.~(19, 20) contracted with terms of
the form $ d_{p n} \de^{(p) A'} $ show that the variation $ \int d^3 x
$ \break $ \[ \de ( \ln \Psi ) / \de e^{A A'}_{~~~~i} (x) \] \de e^{A
A'}_{~~~~i} (x) $ can be found by writing the general variation $ \de
e^{A A'}_{~~~~i} (x) $ as $$ \eqalignno {
\de e^{A A'}_{~~~~i} (x) &= 2 e_{B B' i} n^{C B'} \sum_p \sum_n e_{p
n}
\de^{(p) A'} \rga^{(n) A B}_{~~~~~~C} \cr &~~- {8 \over 3} e_{B B' i}
n^{A B'} \sum_p \sum_m f_{p m}
\de^{(p) A'} \beta^{(m) B}~, &(23) \cr } $$ where $ e_{p n} $ and $
f_{p m} $ are even quantities. This variation in $ e^{A A'}_{~~~~i}
(x) $ produces the variation $$ \de \( \ln \Psi \) = {- 16 \over 3
\hbar \kap^2} \int d^3 x h^{1 \over 2} n^A_{~~A'}
\sum_p \sum_m \lam_m f_{p m} \de^{(p) A'} \beta^{(m)}_A~. \eqno (24)
$$

Thus $ \de (\ln \Psi) $ is determined uniquely for a typical variation
$ \de e^{A A'}_{~~~~i} (x) $, through the $ \ol S_{A'} \Psi = 0 $
constraint (10).  An analogous differential equation for a
finite-dimensional system is $ \pt f / \pt \ul x = \ul F (\ul x) $,
which has a unique solution given a starting value $ f \( \ul x_0 \)
$. Similarly, the $ \ol S_{A'} \Psi = 0 $ constraint in the bosonic
sector must have a unique solution $ \Psi \bigl ( e^{A A'}_{~~~~i} (x)
\bigr ) $, up to a constant factor, related to a starting value $ \Psi
\bigl ( e^{A A'}_{(0) i} (x) \bigr ) $, where $ e^{A A'}_{(0) i} (x) $
is a reference spatial tetrad.

The wave function $ \Psi $ will have a semi-classical expansion $$
\Psi \sim \( A_0 + \hbar A_1 + \hbar^2 A_2 + \ldots \) \exp \( - I /
\hbar \)~, \eqno (25) $$ where $ I $ is a certain action functional of
the spatial tetrad, and $ A_0, A_1, A_2, \ldots $ are loop prefactors.
Inserting this into the $ \ol S_{A'}
\Psi = 0 $ constraint (5), one obtains $$ \eps^{i j k} e_{A A'
i}^{~~~~~3 s} D_j \psi^A_{~~k} + \half \kap^2
\psi^A_{~~i} {\de I \over \de e^{A A'}_{~~~~i}} = 0~. \eqno (26) $$
The uniqueness of the solution of the constraint (10) then implies
that $$ \Psi = c_0 \exp \( - I / \hbar \)~. \eqno (27) $$
[Alternatively, one can examine the constraint (10) for $ \hbar \ln
\Psi $.  This differential equation has no $ \hbar $ in it, and so one
expects that the solution has the form $ \hbar \ln \Psi = - I + $
const.] Here $ \Psi $ obeys the quantum $ S^A $ and $ \ol S^{A'} $
constraints. It obeys the Lorentz rotation constraints, since $ I $
must be formed in a Lorentz-invariant way by contraction of all spinor
indices. Hence $ \Psi = c_0 \exp \( - I / \hbar \) $ gives the general
purely bosonic solution of the quantum constraints. The classical
action functional $ I $ depends on the 3-geometry only though the
3-metric $ h_{i j} $, and its Hamilton--Jacobi trajectories gives
solutions of the classical positive-definite Einstein equations.

The above uniqueness refers to a purely bosonic state $ \Psi \( e^{A
A'}_{~~~~i} (x) \) $ containing no fermions. One can also examine the
``filled'' state in which all fermionic states are filled. In the
representation $ \wti \Psi \( e^{A A'}_{~~~~i} (x),~\ti
\psi^{A'}_{~~i} (x)
\) $, this corresponds to the $ \ti \psi^0 $ part of the wave
function. This obeys the $ S_A \wti \Psi = 0 $ constraint (9), which
similarly gives a solution $$ \wti \Psi = c_1 \exp \( I / \hbar \)~,
\eqno (28) $$ where $ I $ is the same action as in Eq.~(27). This
state is unique among the filled states.

Exponential solutions of the type $ {\rm const.} \exp ( \pm I / \hbar)
$ have previously been found in mini-superspace examples, where
supergravity is quantized subject to the Bianchi I or IX Ansatz [3, 4,
5]. It was also found in [4, 5] that no fermionic states were allowed,
because of the restrictive nature of the $ S_A \Psi = 0 $ and $ \ol
S_{A'} \Psi = 0 $ constraints imposed together. Similarly one can show
that there are no fermionic states at levels $ \psi^{2 n} (n = 1, 2,
3, \ldots) $ in the full theory of supergravity studied here. To see
this, note that the supersymmetry constraints (5), (6) each give an
equation for $ \de \Psi / \de e^{A A'}_{~~~~i} (x) $ at level $
\psi^{2 n} $. These equations follow on the lines of Eqs.~(23), (24).
For consistency, these first-order equations must be identical,
otherwise one obtains $ \Psi = 0 $ at order $ \psi^{2 n} $. But if the
two supersymmetry constraints $ \ol S_{A'} \Psi = 0,~S_A \Psi = 0 $
give identical equations for $ \de \Psi / \de e^{A A'}_{~~~~i} (x) $
at order $
\psi^{2 n} $, then the form of their anti-commutator contradicts the
form of the Hamiltonian constraint $ {\cal H}_{A A'} = 0 $ at the
corresponding order. Thus there are no fermionic states at levels $
\psi^{2 n} (n = 1, 2, 3, \ldots) $. The case with an infinite number
of fermions might of course be different.

Consider now the case in which the 3-surface has the topology of $ S^3
$. On general grounds, there are two preferred quantum states in this
case - the Hartle--Hawking state [6] and the wormhole ground state
[7]. The Hartle--Hawking state is defined by a path integral in which
one fills in inside the 3-surface subject to the given data on it; the
wormhole ground state is defined by a path integral in which one sums
over fields outside the 3-surface which are asymptotically Euclidean,
i.e.~subject to asymptotic flatness at infinity. Since for $ N = 1 $
supergravity we have only two allowed quantum states, there should be
a simple linear relation between these and the Hartle--Hawking and
wormhole ground states. This relation can be checked by studying the
exact Friedmann case of spherical symmetry with radius $ a $ of the
3-sphere [8]. One finds the wormhole state $ \exp \( - 3 a^2 / \hbar
\) $ for the bosonic state $ \exp \( - I / \hbar \) $, and the
Hartle--Hawking state $ \exp \( 3 a^2 / \hbar \) $ for the filled
state $
\exp \( I / \hbar \) $. Hence quite generally (on $ S^3 $) $ c_0 \exp
\( - I /
\hbar \) $ gives the wormhole ground state, and $ c_1 \exp \( I /
\hbar \) $ gives the Hartle--Hawking state. It would be of interest to
investigate this relation for other compact topologies.

Making a loop expansion of the wormhole state, one has $$ c_0 \exp \(
- I / \hbar \) \sim \( B_0 + \hbar B_1 + \hbar^2 B_2 + \ldots
\) \exp \( - I_1 / \hbar \)~, \eqno (29) $$ where $ I_1 $ is the
classical action outside the $ S^3 $ subject to asymptotic flatness
and to the prescribed 3-metric $ h_{i j} $ on the $ S^3 $. Hence $ I =
I_1 $, and $ c_0 \exp \( - I / \hbar \) $ is the wave function for the
wormhole state. Similarly $ c_1 \exp \( I / \hbar \) $ is the
Hartle--Hawking wave function, where $ - I $ is the classical action
found by solving the positive- definite Einstein equations inside the
$ S^3 $ with prescribed 3-metric $ h_{i j} $ on the $ S^3 $. This
shows that there is a connection between the dynamical field equations
of supergravity and the initial or boundary conditions for the wave
function.

This highly restrictive form of the solution to the quantum
constraints may be a feature only of $ N = 1 $ supergravity, or
perhaps also of higher - $N$ supergravity without couplings. When one
turns to supergravity coupled to supermatter, it is immediate that
many more solutions to the constraints are possible, essentially
because the wave function $ \Psi $ depends on more fields, but does
not have to obey any more constraints than in pure supergravity. The
special form of the solution found here should be a feature of pure
supergravity only.

It will, of course, also be of interest to compare the approach of
this paper with the standard approach based on scattering theory, to
understand why scattering states should be forbidden.  It is now
clear, however, that $ N = 1 $ supergravity is a theory without
quantum ultraviolet divergences, and without any quantum corrections
at all.
\vskip 18 pt
\ni {\bf References}

\ni [1] P.D. D'Eath, {\it Phys.~Rev.~D} {\bf 29}, 2199 (1984).

\ni [2] C. Teitelboim, {\it Phys.~Rev.~Lett.} {\bf 38}, 1106 (1977).

\ni [3] R. Graham, {\it Phys.~Rev.~Lett.} {\bf 67}, 1381 (1991).

\ni [4] P.D. D'Eath, S.W. Hawking and O. Obreg\'on, {\it
Phys.~Lett.~B} {\bf 300}, 44 (1993).

\ni [5] P.D. D'Eath, submitted to {\it Phys.~Rev.~D.}

\ni [6] J.B. Hartle and S.W. Hawking, {\it Phys.~Rev.~D} {\bf 28},
2960 (1983).

\ni [7] S.W. Hawking and D.N. Page, {\it Phys.~Rev.~D} {\bf 42}, 2655
(1990).

\ni [8] P.D. D'Eath and D.I. Hughes, {\it Phys.~Lett.~B} {\bf 214},
498 (1988); {\it Nucl.~Phys.~B} {\bf 378}, 381 (1992.

\bye